\newcommand\apj{{ApJ\,}}%
\newcommand\apjl{{ApJ\,}}%
\newcommand\apjs{{ApJS\,}}%
\newcommand\apss{{Ap\&SS\,}}%
\newcommand\aap{{A\&A\,}}%
\newcommand{\Iso}[2]{^{#1}{\rm #2}}
\documentclass[graybox, natbib, footinfo]{svmult}

\usepackage{mathptmx}       
\usepackage{helvet}         
\usepackage{courier}        
\usepackage{type1cm}        
\usepackage{makeidx}         
\usepackage{graphicx}        
\usepackage{multicol}        
\usepackage[bottom]{footmisc}
\usepackage{url}
\makeindex             

\begin{document}
\title*{Helium burning in moderate-mass stars}
\author{Achim Weiss}
\institute{A. Weiss\at Max-Planck-Institut f\"ur Astrophysik,
  Karl-Schwarzschild-Str. 1, 85748 Garching, Germany, \\
  \email{aweiss@mpa-garching.mpg.de}}

\maketitle

\abstract{The evolution of low- and intermediate mass stars at the
  onset and during core helium burning is reviewed. Particular
  emphasis is laid on structural differences, which may allow to
  identify a star's nature and evolutionary phase in spite of the fact
that it is found in a region of the Hertzsprung-Russell-Diagram
objects from both mass ranges may populate. Seismic diagnostics which
are sensitive to the temperature and density profile at the border of
the helium core and outside of it may be the most promising tool.}

\section{Mass ranges}
\label{weiss_s:1}

In this talk I will be concerned with stellar models in the mass
ranges usually called {\em low} and {\em intermediate}. While these
two ranges are separated by a physical effect -- the ignition of
helium under degenerate plasma conditions -- stars in both ranges
share some properties, but within each mass range they also differ in
others. Here, we are concentrating on stars which 
\begin{itemize} 
\item are massive enough to ignite helium and had the time to reach
  the helium-burning stage within a Hubble-time, and
\item at helium ignition are not brighter than the tip of the red
  giant branch, i.e.\ have approximately $\log L/L_\odot <  3.5$.
\end{itemize}

These requirements limit the mass range of interest to about $0.7 <
M/M_\odot < 7$; exact numbers, which are not needed here, would depend
on composition. I call this mass range that of {\em moderate-mass}
stars.

\begin{figure}
\centerline{\includegraphics[scale=0.55]{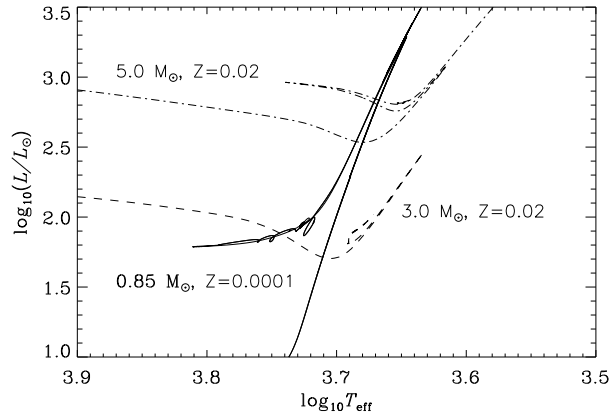}}
\caption{Parts of the evolutionary tracks in the
  Hertzsprung-Russell-diagram for a low-mass ($0.85\,M_\odot$) star
  with $Z=0.0001$ (solid line), and two stars with solar-like metallicity
  ($Z=0.02$) and 3 (dashed line) and $5\,M_\odot$ (dash-dotted line)}
\label{weiss_f:1}    
\end{figure}

Before and during helium burning moderate-mass stars may be confused,
as is demonstrated by the example of Fig.~\ref{weiss_f:1}, which shows
stars of 0.85, 3.0, and 5.0~$M_\odot$ before and during core helium
burning. If the metallicity is not known accurately enough, a low-mass
star (LMS) on its ascent on the RGB occupies roughly the same effective
temperature and luminosity as a more metal-rich intermediate-mass
star (IMS) of $3\, M_\odot$ during core helium burning. Close to the tip of
the RGB it crosses the evolution of a $5\, M_\odot$ star before and
shortly after  helium ignition  and again during early helium shell
burning. In the following I will discuss the structural differences
between such models, which occupy similar regions in the HRD.

\section{Low-mass stars: the core helium flash}
\label{weiss_s:2}

All of the following is based on traditional 1-dimensional,
hydrostatic stellar evolution models, which have been calculated
through their full evolution from the main-sequence into and past core
helium burning; in particular the core helium flash is followed
completely, which is possible with the GARSTEC code \citep{wsch:2008}
which I used. One should keep in mind that these models are still
approximations to the real evolution, which is dynamic and
multi-dimensional. Such models will be presented by M.~Mocak (these
proceedings). 

The stability of nuclear burning depends mainly on the thermodynamic
properties of the plasma, as can be shown with very simple arguments
\cite[following][]{kw:90}. If we denote a small perturbation to physical
quantities by an index 1, we can write the equation of state in the
following standard way
\begin{equation}
\frac{\rho_1}{\rho} = \alpha\frac{P_1}{P} -\delta\frac{T_1}{T}
\label{weiss_e:1}
\end{equation}
with $\alpha$ and $\delta$ being the usual logarithmic derivatives
with respect to $P$ and $T$ of density. For the energy generation rate
we assume a power law, $\epsilon = \epsilon_0\rho^n T^\nu$, and define
\begin{equation} 
\xi=\frac{4\delta}{4\alpha-3}, \qquad \eta=\alpha\xi-\delta =
\frac{3\delta}{4\alpha-3}.
\label{weiss_e:2}
\end{equation}
With
\begin{equation}
A = \frac{\epsilon_0(n \eta + \nu)}{c_P T_0 (1-\nabla_\mathrm{ad} \xi)}
\label{weiss_e:3}
\end{equation}
one can show that $T_1/T \sim \exp(At)$, i.e., if $A<0$, nuclear
burning is stable, while for $A>0$ 
it is unstable, and an exponential runaway will take place for small
perturbations. 

For a main-sequence star with almost ideal gas conditions,
$\alpha=\delta=1$, $n=1$, and thus $\xi=4$ and $\eta=3$, such that
$A<0$. However, for a LMS near helium ignition, the helium
electron gas is (non-relativistically) degenerate and dominating the equation of
state. Therefore, $\alpha=3/5$ and $\delta \ll 1$. $\xi$ and $\eta$
are thus $\approx 0$ and $A>0$; helium ignition is thermally
unstable and a runaway takes place. This is the reason for the {\em
  core helium flash}.

\begin{figure} 
\hspace*{-0.3cm}
\includegraphics[scale=0.37]{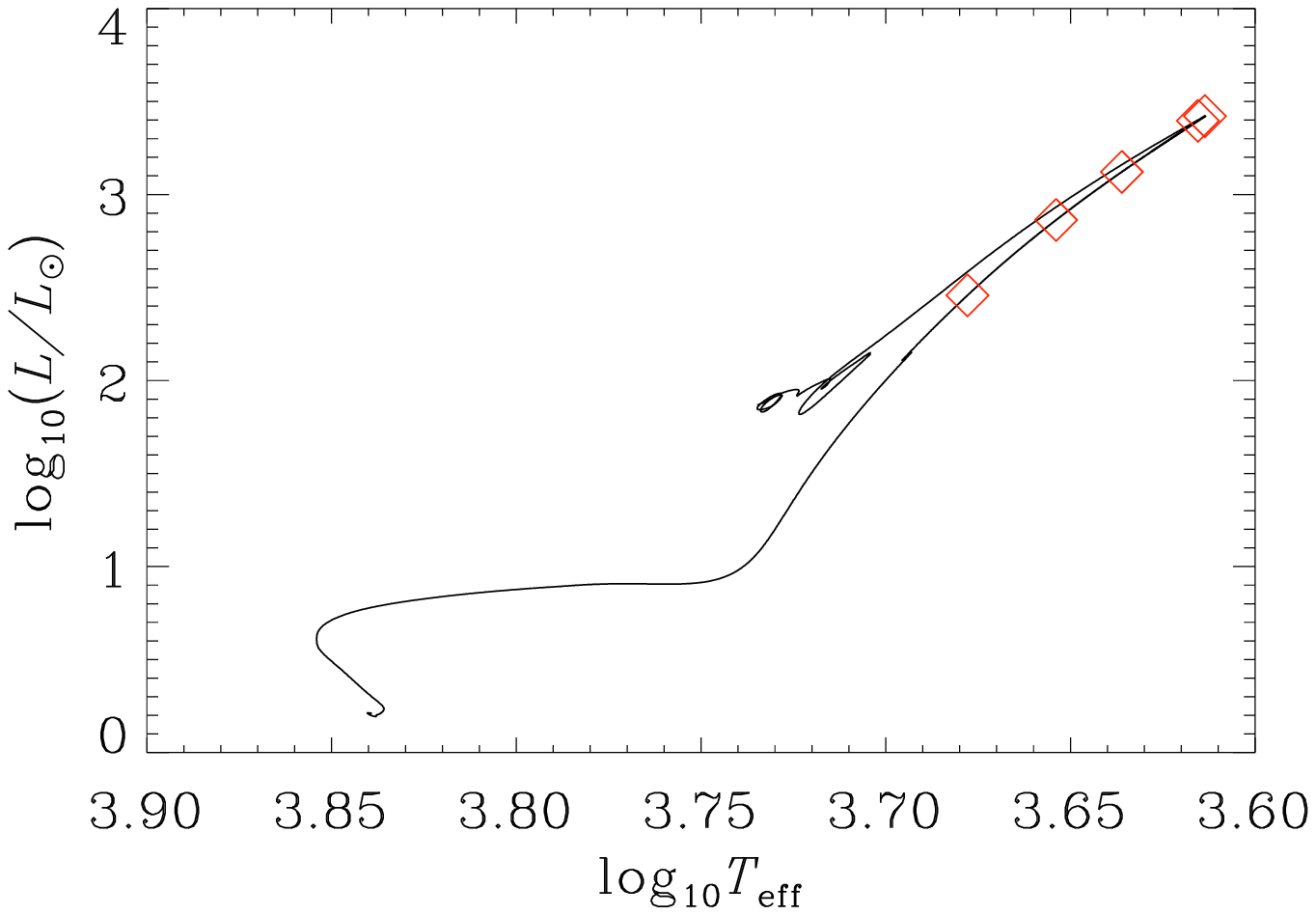}
\hspace*{-0.3cm}
\includegraphics[scale=0.18]{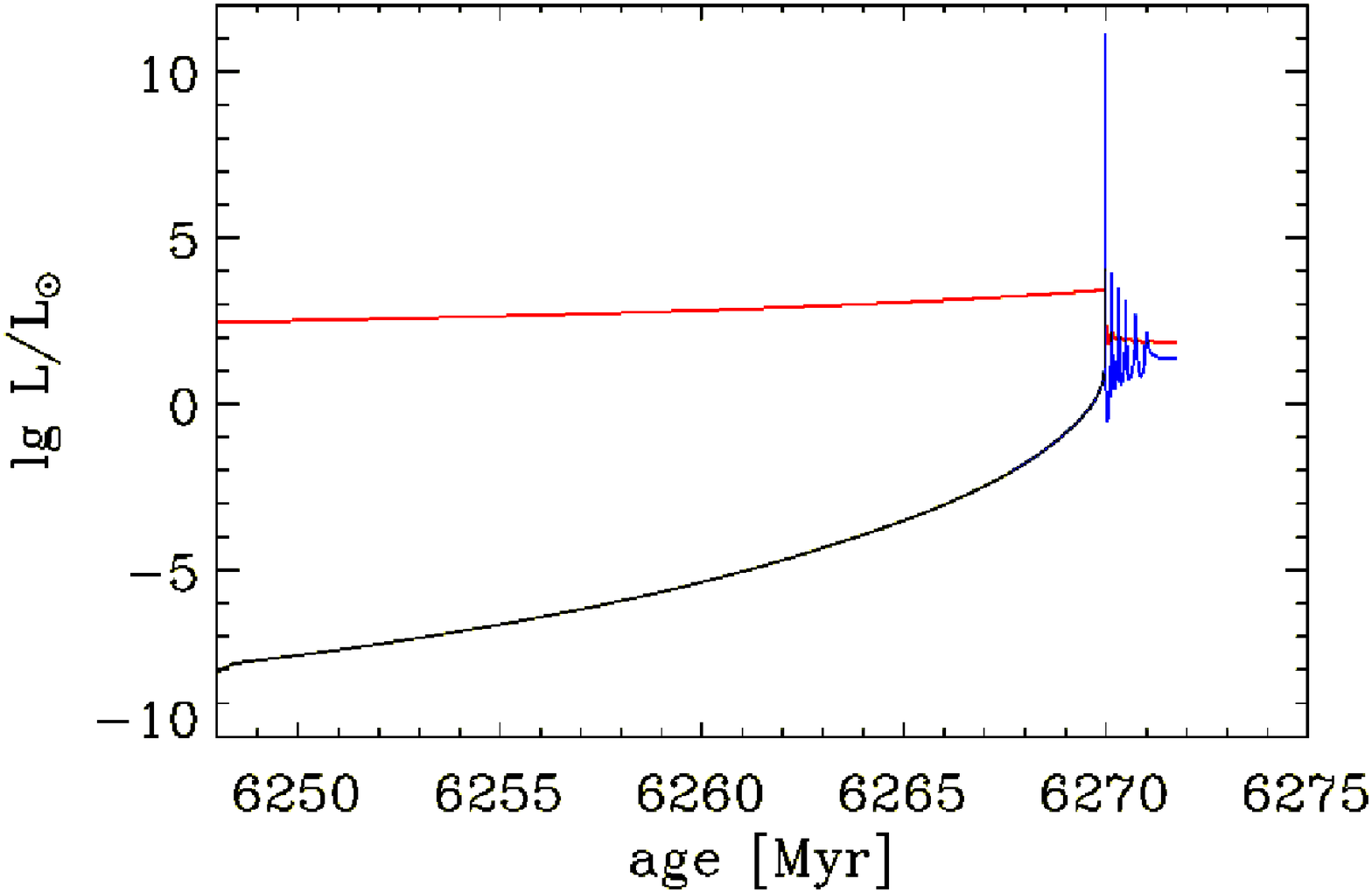}
\caption{Left: Evolution of a star of $1\,M_\odot$ star with $Z=0.001$ from
  the main-sequence through the core helium flash to the horizontal
  branch. Along the RGB the diamonds indicate points where the helium
  luminosity $L_\mathrm{He}/L_\odot$ has reached levels of
  $10^{-8}$, $10^{-5}$, $10^{-3}$,  $10^{0}$, and $10^{10}\,
  L_\odot$. The last two points are almost at the same location.
  Right: The rise of $L_\mathrm{He}$ (blue line) with time in
  comparison with the almost constant total luminosity $L$ during the
  final stages of the helium flash and the approach to the HB }
\label{weiss_f:2}
\end{figure}

In spite of its name, the flash starts very slowly and quite early on
the RGB. Figure~\ref{weiss_f:2} demonstrates how the helium luminosity
increases from values of $10^{-8}\, L_\odot$ at $\log L/L_\odot
= 2.45$ to peak values of $10^{10}\, L_\odot$ (and above) at the
tip. While the star needs  18~Myr from the first to the second point ($\log
L_\mathrm{He}/L_\odot = -3$) at $\log L/L_\odot = 3.12$, already
3.3~Myr later $\log L_\mathrm{He}/L_\odot = 3$ is reached (basically
at the tip), and a mere 180~yrs later the peak luminosity of $10^{10}
\,L_\odot$ is reached (the two final diamond symbols in
Fig.~\ref{weiss_f:2} are at almost the same position). In this final
phase the exponential runaway truly takes place.

The right panel shows how $\log L_\mathrm{He}/L_\odot$ increases
during the final stages of the helium flash in comparison to the
almost constant total luminosity. The approach to the horizontal
branch (HB) is characterized by secondary flashes in the course of the
heating of the core and the progression of helium burning from the
off-centre ignition shell to the centre. These correspond to the small
loops visible in the left plot of the same figure. 

\begin{figure} 
\centerline{\includegraphics[scale=0.3]{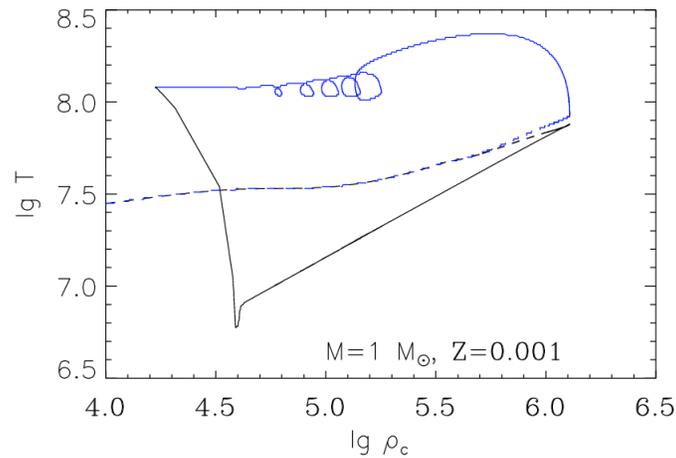}}
\caption{Evolution of central ($T_c$; black, lower curve) and maximum
  ($T_\mathrm{max}$; blue, upper curve) 
  density and temperature during the core 
  helium flash in the same model as in Fig.~\ref{weiss_f:2}. The
  evolution proceeds first from left to right, where both curves
  separate during the flash, then move to the left as the core is
  expanding and finally meet again after the secondary flashes are over}
\label{weiss_f:3}
\end{figure}

The ignition of helium happens off-centre because of three combined
effects: (i) the helium core being inert, to first order it is
isothermal, and its temperature defined by the hydrogen shell
surrounding it. At the tip of the RGB, the shell temperature is $\log
T \approx 7.9$; (ii) as the core is growing in mass due to the
progression of the hydrogen shell, thermal energy is released, since
the core, being degenerate, is shrinking in radius and thereby matter
is compressed; this creates a positive $T$-gradient towards the
centre; (iii) emission of plasma neutrinos is increasingly cooling the
core, with two effects: the maximum temperature in the core is limited
and helium ignition delayed, and the centre is cooler than outer
regions of the core. In consequence, the temperature maximum is at
some place outside the centre (typically around a relative mass of
0.1\ldots 0.2), but detached from the hydrogen shell. For higher
masses, the core being not degenerate, $T$ is increasing towards the
centre, and helium is ignited centrally, and at much lower helium core
mass because of the lack of efficient neutrino cooling. The evolution
of density and temperature at the centre and at the location of
maximum temperature during the core helium flash is shown in
Fig.~\ref{weiss_f:3}.

The evolution along the RGB is, for LMSs, determined by the
mass of the helium core. Simple homology considerations lead to the
global relations 
\begin{equation} 
L \sim M_c^7; \qquad T \sim M_c; \qquad L \sim \mu^7.
\label{weiss_e:4}
\end{equation}
More detailed considerations, which include $M$--$R$--relations for
degenerate structures lead to slightly different exponents. 
These qualitative relations are confirmed by numerical models. The
consequence of (\ref{weiss_e:4}) is that independent of total mass,
all LMSs reach helium ignition temperatures at virtually the
same core mass and therefore identical luminosities. This is the
reason that the RGB-tip can serve as a standard candle. 

On closer inspection, $M_c$ and $L$ at the RGB-tip do depend slightly
on stellar parameters, but also on the input physics, and even on the
numerical code used. \cite{scbook} give the following gross
dependencies: (i) the core helium mass decreases with increasing
metallicity $Z$ by about $0.001-0.01\,M_\odot$, and with increasing total mass
$M$ by $0.01\,M_\odot$ (taking a typical range in $Z$ and $M$ for old
halo stars); (ii) the luminosity $\log L/L_\odot$ increases with $Z$ by
about 0.1~dex and to a lower degree with decreasing mass ($0.05$~dex).
\cite{rw:92} give analytical fits to their numerical results:
\begin{eqnarray}
 M_{c} &=& 0.475 - 0.22 (Y_e -0.25) -0.01 (3- \log Z) -0.025
 (M/M_\odot-0.8) \nonumber \\ 
M_\mathrm{bol,tip} &=& -3^{m}.27 +1.1 (Y_e-0.25) -0.21 (3- \log Z) +0.30
(M/M_\odot-0.8),
\label{weiss_e:5}
\end{eqnarray}
($Y_e$ being the helium mass fraction in the envelope). These
relations include the effect of mass loss along the RGB.

Recently, Cassisi (Workshop on {\em The Giant Branches}; Lorentz
Center, Leiden, The Netherlands; 2009; see {\tt www.lorentzcenter.nl})
showed model comparisons 
between different codes. The bolometric tip brightness differed by
about 0.2~mag for a wide range of metallicities. With the input
physics predefined, differences are about half as large, as Weiss
(same meeting) demonstrated. Nevertheless, they are at least as
important as the global stellar parameters mentioned above.

Updates in input physics have also been the reason for changes in the
core mass and luminosity at the RGB-tip.
\cite{cppcs:07} investigated the influence of new conductive opacities
and found an almost metallicity-independent decrease in $\log
L/L_\odot$ of 0.03~dex for the TRGB, and 0.02~dex for the ZAHB
(zero-age HB), when using the new opacities. New nuclear rates for the
$3\alpha$ and $\Iso{14}{N}(p,\gamma)\Iso{15}{O}$ reactions were tested
by \cite{wsksch:05}, both rates being now lower than older values. In
the first case, luminosity increases by a small amount of $\approx
0.01$~dex (helium igniting later), while in the case of the
CNO-bottleneck reaction it decreases by $\approx 0.05$~dex. This is
one of the largest influences on tip quantities found so far. 

An open question concerns the possibility of significant mixing
episodes besides that by the convective envelope. C.~Charbonnel
reports about thermohaline mixing between the lower boundary 
of the convective envelope and the outer regions of the hydrogen
burning shell, and M.~Mocak (both these proceedings) about the
possibility of mixing between core and envelope during the helium
flash. So far, such mixing episodes were found only in extreme
situations: either in metal-free LMSs \citep[e.g.\ ][]{sscw:02}
or in ordinary Pop.~II stars with extremely thin envelopes
\citep{cssw:03}. 

After the core helium flash the next phase of low-mass stellar
evolution is the (zero-age) horizontal branch, where the core helium
burning takes place. The ZAHB luminosity is, within the errors, in
agreement with observations, but differs between codes and depending
on the implemented physics by about 0.05 to 0.1~dex in $\log
L/L_\odot$ (Cassisi, {\em The Giant Branches}, 2009). 

\begin{figure} 
\includegraphics[scale=0.30,bb= 1 1 610 580]{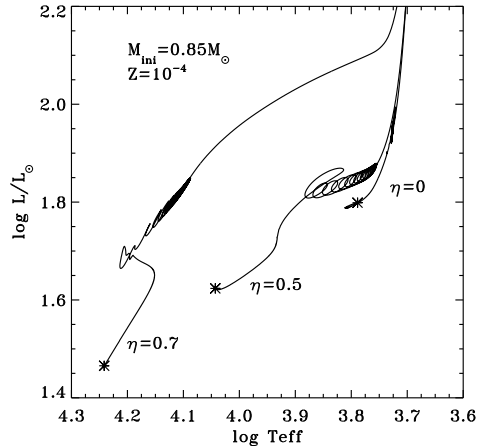}\hfill
\begin{minipage}[t]{0.35\linewidth}
\vspace*{-4.cm}
\caption{Horizontal branch evolution for three stellar models
  (initial mass $M=0.85\,M_\odot$; $Z=0.0001$), which were followed
up the RGB and through the core helium flash with different values of
$\eta$, the parameter in the Reimers mass loss formula. The tracks
start at the ZAHB, indicated by the asterisk.}
\label{weiss_f:4}
\end{minipage}\hfill
\end{figure}

The stars burn helium in a convective core, and evolve, with
increasing central CO-abundance and thus molecular weight, to higher
luminosities. Whether $T_\mathrm{eff}$ in- or decreases initially,
depends on the contribution of H- and He-burning to the total
luminosity. At core helium exhaustion the tracks always turn 
towards cooler temperatures and up to start the early AGB evolution,
unless the envelope mass is extremely low. I show three cases in
Fig.~\ref{weiss_f:4}, taken from \cite{saw:05}, which are the result
of full flash calculations with different degrees of mass loss
(according to the Reimers formula, characterized by the parameter
$\eta$) along the RGB. With increasing mass loss, the envelope mass on
the HB decreases and the stars settle down at increasingly hotter
$T_\mathrm{eff}$. The cases shown have HB masses of 0.847, 0.657, and
$0.563\, M_\odot$, but, as we saw above, the same core
mass. Complications arise due to the retreating 
convective core, which leaves behind a region of varying molecular
weight, which gives rise to semiconvection, an ill-understood secular
mixing process. \cite{scbook} describe a method how to treat this in
stellar evolution programs. In the case shown in Fig.~\ref{weiss_f:4},
standard convection and the Schwarzschild-criterion for convective
stability was used. This treatment leads to so-called {\em breathing
  pulses}, which are episodes of convection outside the core that
result in mixing of fresh helium into the core and the loops visible
in the figure. These are most likely artificial effects which
disappear, for example, when the Ledoux-criterion for convection is
used, since this tends to inhibit convection.

\begin{figure} 
\hspace*{-0.3cm}
\includegraphics[scale=0.35]{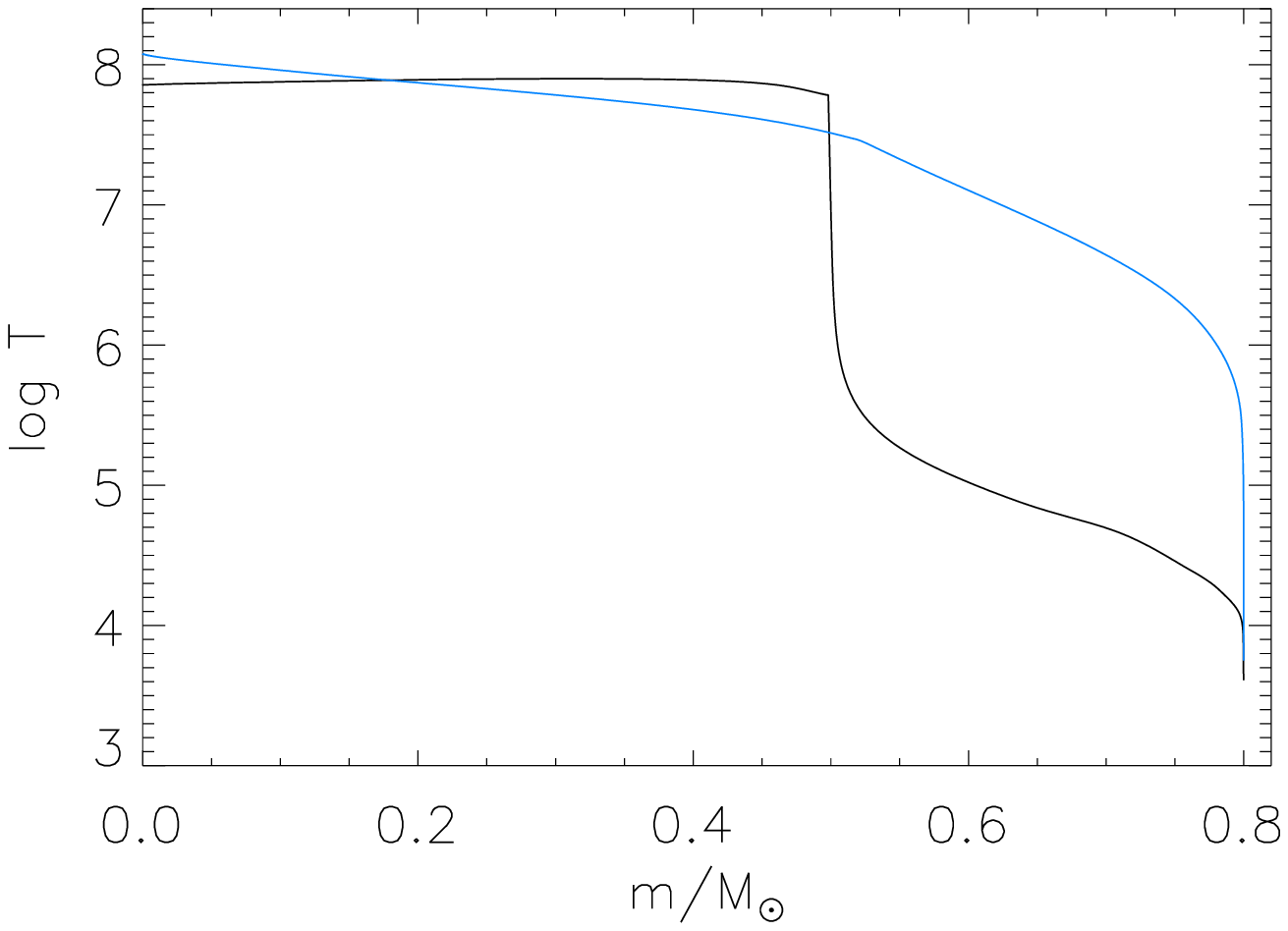}
\hspace*{-0.3cm}
\includegraphics[scale=0.35]{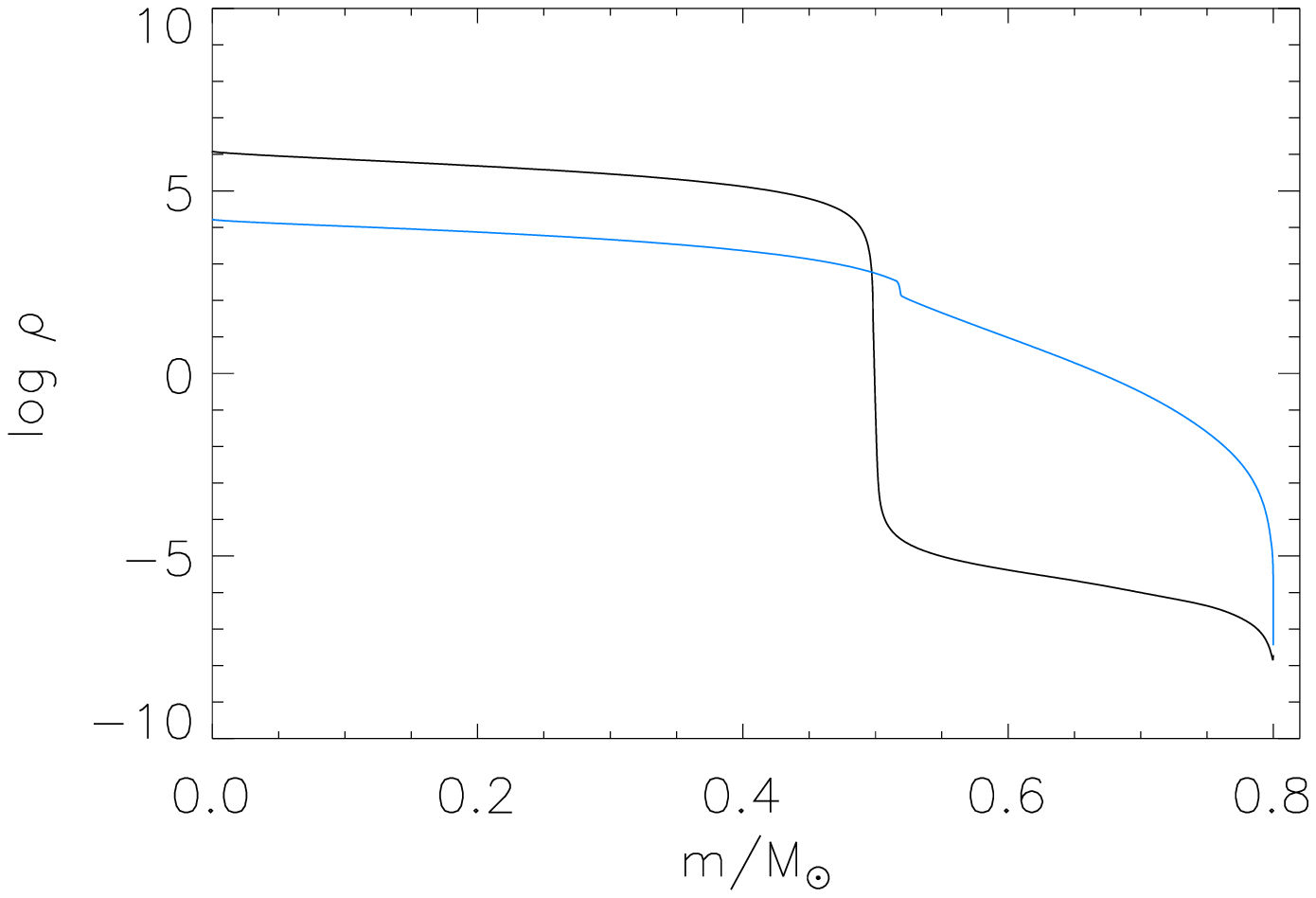}
\caption{Temperature (left) and density (right) profiles in a star of
  $0.8\,M_\odot$ and $Z=0.001$ before (black) and after (blue) the
  core helium flash.}
\label{weiss_f:5}
\end{figure}

The chemical profile hardly changes from the pre- to the post-flash
phase. The helium core increases by at most a few 1/100~$M_\odot$,
with a very steep composition gradient being kept. The central carbon
abundance increases to levels of 3-5\% due to the $3\alpha$-reactions
taking place during the $\approx 10^6$~yrs it takes the star to settle
down on the ZAHB. Stellar models, which are not followed through the
flash, but where the calculations are stopped at the RGB-tip and
resumed on the ZAHB, need to change the chemical composition in this
respect to be reasonable approximations to full flash calculations
\citep{saw:05}. 

Since the main effect of the core helium flash is the lifting of
degeneracy in the core by expansion and
heating (see Fig.~\ref{weiss_f:3}), the $T$- and $\rho$-profiles
before and after ignition differ 
drastically (Fig.~\ref{weiss_f:5}): Both are now monotonically
decreasing and the very steep gradient at the helium core boundary has
vanished almost completely (a small step in density is still
discernible). Together with the change of a radiative to an adiabatic
temperature gradient at the convective core boundary, seismology
should be able to detect these 
differences, and therefore be able to discriminate between pre- and
post-flash structures. 

\section{Intermediate-mass stars: quiescent helium ignition}
\label{weiss_s:3}

\begin{figure}
\includegraphics[scale=0.4,bb=70 70 500 350]{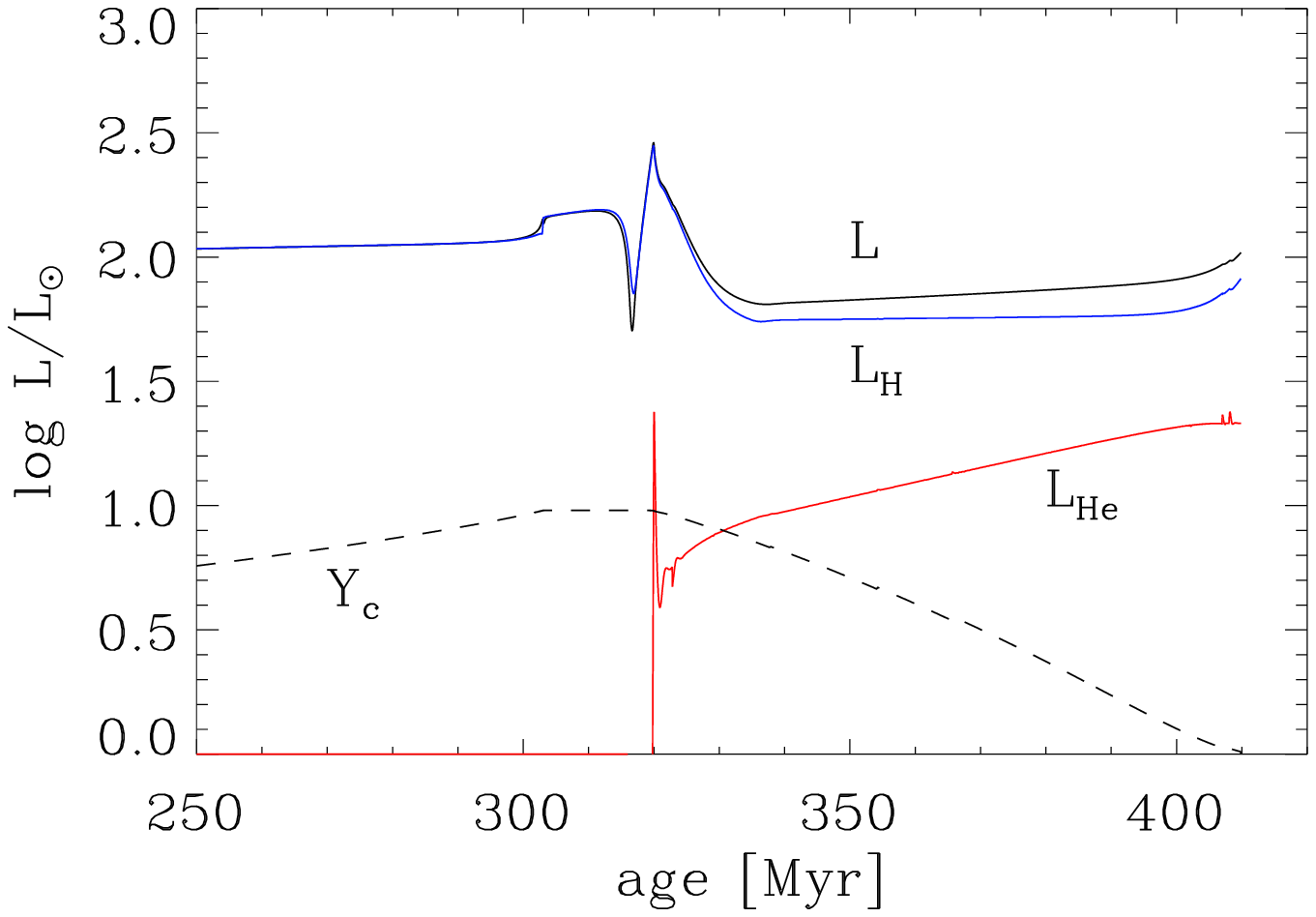}
\hspace*{-0.3cm}
\includegraphics[scale=0.4,bb=70 70 500 350]{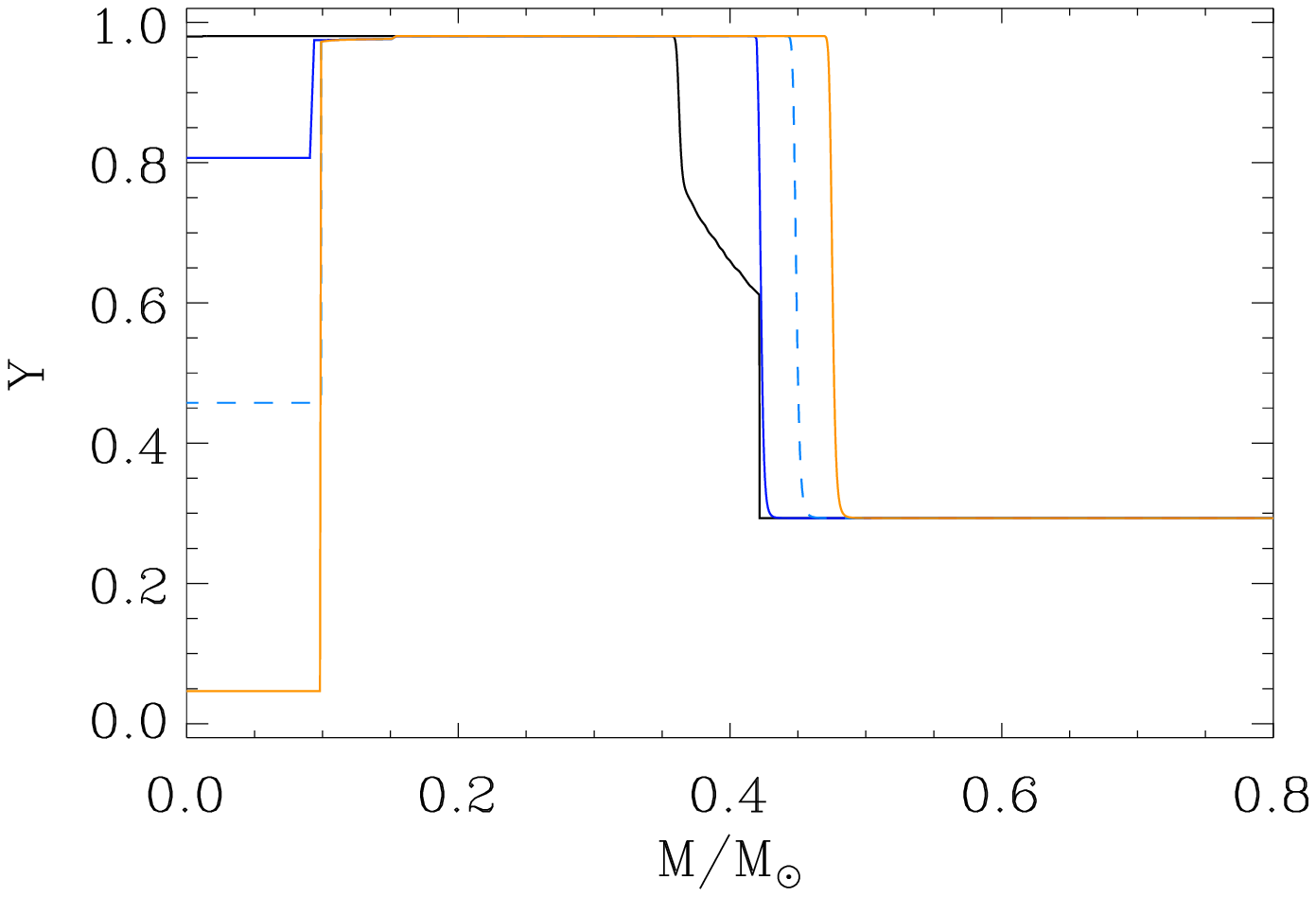}
\caption{Left: Total, hydrogen- and helium-luminosity during the ignition of
core helium burning in a $3\,M_\odot$ star ($Z=0.02$). The central
helium content $Y_c$ is also shown. Right: Helium abundance inside the
core of the same model for four different stages during core helium burning}
\label{weiss_f:6}
\end{figure}

Figure~\ref{weiss_f:6} (left panel) displays the evolution of
luminosities during the helium ignition in a $3\,M_\odot$ star. In
contrast to LMSs, $L_\mathrm{He}$ rises only to moderate
levels, which, after some adjustment phase are of the same order as
those during the burning phase. The decrease in total luminosity $L$,
which is visible in the HRD of Fig.~\ref{weiss_f:1}, is solely due to
the reduction of the luminosity produced in the H-shell. The rise of
$\log L_\mathrm{He}/L_\odot$ from -2 to 0.6 during ignition takes
2~Myrs in this model. If one compares helium core sizes and
luminosities during core helium burning between the $0.85$ and
$3\,M_\odot$ stars of Fig.~\ref{weiss_f:1}, one finds that they differ
by only $0.13\, M_\odot$ resp.\ 0.8~dex. The general run of these
quantities with mass can be found in Fig.~5.19 of \cite{scbook}:
Almost independent of metallicity $M_c$ it is nearly constant for $M < 2
M_\odot$ (as discussed in Sect.~\ref{weiss_s:2}), then drops over a
rather narrow mass interval to values of about $0.3\,M_\odot$. The
location and width of the drop depends on metallicity. With increasing
mass it then increases almost linearly, reaching the low-mass values
again around $4\,M_\odot$. Luminosity behaves in a very similar way,
but low-mass values are reached again only at much higher mass. This
is the reason why at the brightness level of the upper RGB only
LMSs are to be expected.

The right panel of Fig.~\ref{weiss_f:6} shows the helium profile in
the inner $0.8\,M_\odot$ 
of the same model for four stages (lines from top to bottom at left
edge: black solid, blue solid, cyan dashed, orange solid): before
ignition ($Y_c=0.98$), at $Y_c=0.804$ and 0.458, and finally close to
exhaustion ($Y_c=0.047$). The burning core is slightly growing in mass
(the calculations were done without overshooting and with the
Schwarzschild criterion), while the hydrogen shell is advancing by
$0.12\,M_\odot$ and steepening the chemical profile. In the HRD these
phases are at the tip of the red giant phase, during the lower end of
the short loop, and on the second red giant ascent. 

\begin{figure}
\includegraphics[scale=0.4,bb=70 70 500 350]{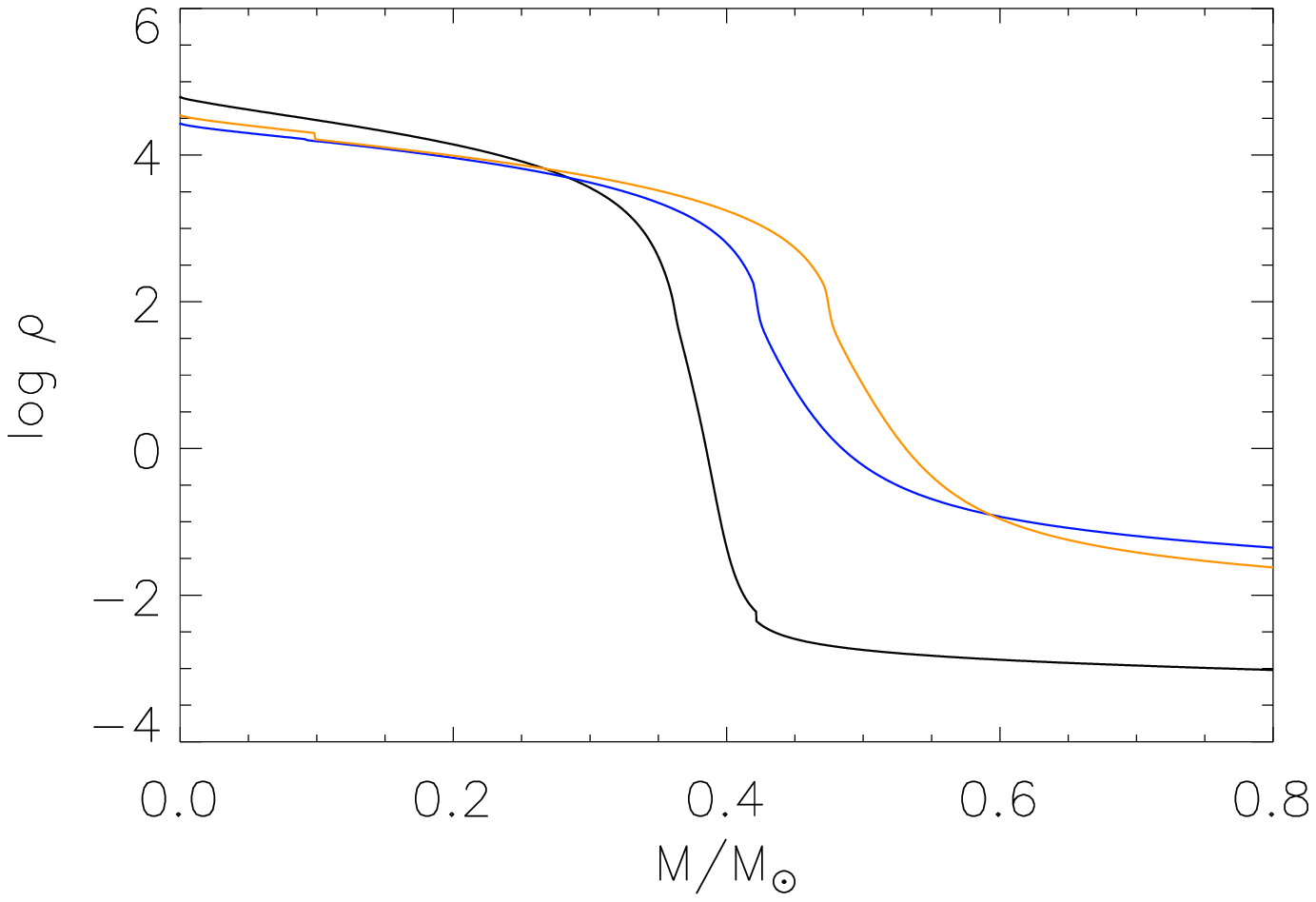}
\includegraphics[scale=0.4,bb=70 70 500 350]{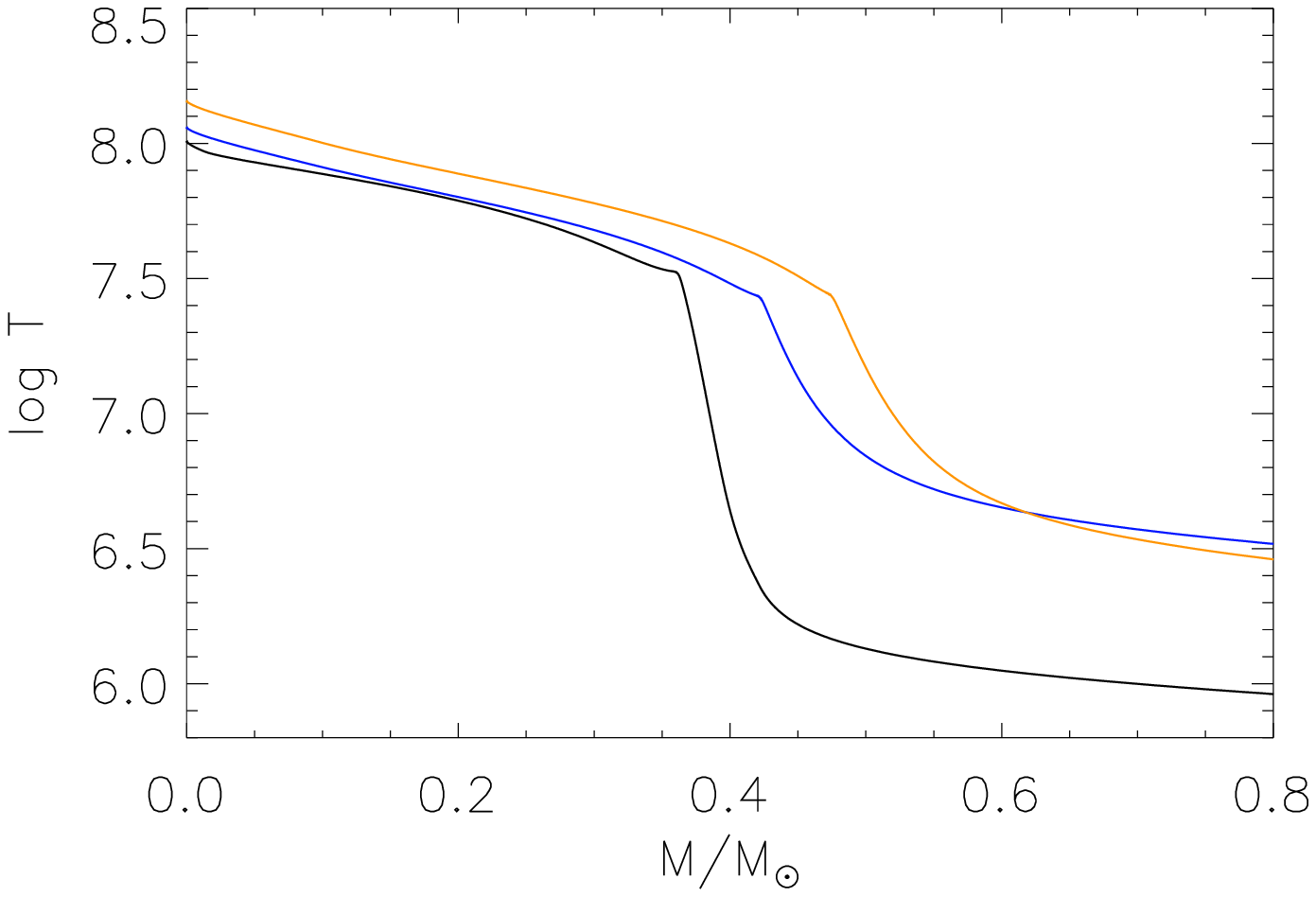}
\caption{Temperature (left) and density (right) profiles within the
  same star as in Fig.~\ref{weiss_f:6} before (at the tip of the
  first RGB ascent; black line), during the early (blue) and the very
  late (red) phases of core helium burning}
\label{weiss_f:7}
\end{figure}

The temperature and density structure of the same model before, during
and at the end of core helium burning is shown in
Fig.~\ref{weiss_f:7}, and should be compared to
Fig.~\ref{weiss_f:5}. The change is not as drastic as in the LMS, and
the pre-ignition drop is much less pronounced. During and 
after core helium burning the profiles are more similar, but the
curvature is different, and the IMS has a broader
hydrogen shell. Note that in these figures the same mass range in
absolute mass is shown. The outer parts of the $4\,M_\odot$ star are not
depicted. 

A particular feature of IMSs during core helium
burning are the {\em blue loops}, during which they cross the Cepheid
intstability strip. These are very sensitive to the
detailed structure of the star, in particular of the chemical profile,
which is established during the main-sequence
\citep{kw:90}. Differences in the appearance and extension of the loops
arise from the treatment of overshooting, semiconvection and
rotation.

\begin{figure} 
\includegraphics[scale=0.4,angle=90]{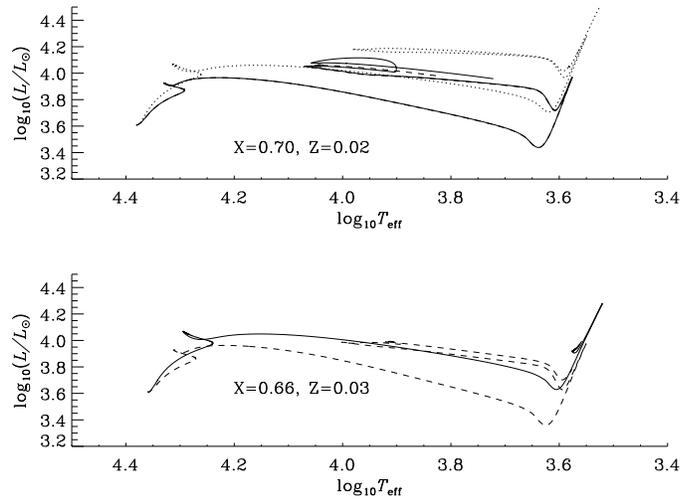}
\caption{A star of $9\,M_\odot$ and two different chemical
  compositions followed from
  the zero-age main sequence until the end of core helium burning with
  different treatments of convection.
  Upper panel ($Z=0.02$): the solid and dashed line are without
  overshooting, and a reduced spatial resolution in the latter case,
  while the dotted line is with overshooting included.  
  Lower panel ($Z=0.03$): without (dashed) and with (solid) overshooting  }
\label{weiss_f:8}
\end{figure}

Fig.~\ref{weiss_f:8} demonstrates this for a $9\,M_\odot$ star and two
different chemical compositions. Cases with overshooting included have
generally higher luminosities during helium burning (as is the case
also during the MS phase), but the loops tend to be reduced, which is
quite obvious in the $X=0.66$, $Z=0.03$ case. Ignoring overshooting
and employing the Schwarzschild criterion for convective instability
it may happen that during the final phases of core helium burning
regions outside the convective core become convective, too, and fresh
helium is mixed into the core, rejuvenating the core and leading to a
secondary loop in the HRD. There is an indication of this along the
dashed line in the $Z=0.03$ case, but it is quite evident in the
solid-line track of the upper panel. Here, the ``rejuvenation'' was
provoked by a reduced spatial resolution in the calculation. Notice
that apart from this phase during the loop the two tracks with higher
and lower numerical resolutions are indistinguishable. A general
problem in up-to-date models is that below $\approx 4\, M_\odot$ they
do not display blue loops (see Fig.~\ref{weiss_f:1}), while observed
Cepheids, which can be identified with this phase, have luminosities
as low as is appropriate for such masses. There is clearly improvement
in the modeling of IMSs before and during helium
burning.

\section{Summary}
\label{weiss_s:4}
Low- and intermediate-mass stars populate similar regions in the HRD
before, during and after core helium burning. If $T_\mathrm{eff}$ and
composition are accurately known, the LMSs can be identified
as the cooler objects. If these parameters are not known,
low-metallicity pre-flash LMSs may be confused with
helium-burning IMSs of higher metallicity. During
helium burning LMSs on the HB may also be found in the same
HRD-region that is occupied by blue loops of IMSs.

The interior structure at the helium core boundary is clearly
different between stars in these two mass ranges, with LMSs
generally showing much steeper $T$ and $\rho$ gradients. This is
particularly true for pre-flash objects. This opens the possibility
for seismology to discriminate between them. If a LMS and an
IMS occupy a similar position in the HRD, their
masses should be quite different, such that also first-order seismic
scaling laws should be able to discriminate between them. And finally,
the IMS will have a convective core, while the
low-mass pre-flash star will have a degenerate core.

\begin{acknowledgement}
This work was supported by the Cluster of Excellence EXC~153 ``Origin
and Structure of the Universe'', Garching ({\tt
  http://www.universe-cluster.de}). 
\end{acknowledgement}

\end{document}